\documentstyle[aps,prl,preprint]{revtex}
\def\la{\mathrel{\mathpalette\fun <}}
\def\ga{\mathrel{\mathpalette\fun >}}
\def\fun#1#2{\lower3.6pt\vbox{\baselineskip0pt\lineskip.9pt
        \ialign{$\mathsurround=0pt#1\hfill##\hfil$\crcr#2\crcr\sim\crcr}}}
\def\l{\lambda}
\begin{document}
\begin{titlepage}
\vspace*{-62pt}
\begin{flushright}
DART-HEP-93/05\\
August 1993
\end{flushright}

\vspace{0.75in}
\centerline{\bf PSEUDO-STABLE BUBBLES$^{\ast}$}
\vskip 1.0cm
\centerline{Marcelo Gleiser}

\vskip 1.5 cm
\centerline{\it Department of Physics and Astronomy, Dartmouth College,
Hanover, NH 03755}

\vskip 1.5 cm
\centerline{\bf Abstract}
\begin{quote}
{The evolution of spherically symmetric unstable scalar field configurations
(``bubbles'') is examined
for both symmetric (SDWP) and asymmetric (ADWP)
double-well potentials.
Bubbles with initial static energies $E_0\la E_{{\rm crit}}$,
where $E_{{\rm crit}}$ is some critical value,
shrink in a time scale determined by their linear dimension, or ``radius''.
Bubbles with $E_0\ga E_{{\rm crit}}$ evolve
into time-dependent, localized configurations which are
{\it very} long-lived compared to characteristic time-scales in the
models examined.
The stability of these configurations is investigated and possible
applications are
briefly discussed.}

\vspace{48pt}
PACS : 11.10.Lm, 98.80.Cq, 64.60.-i\\
$\ast$ electronic mail: gleiser@peterpan.dartmouth.edu

\end{quote}

\end{titlepage}
\def\mpl{{m_{Pl}}}
\def\x{{\bf x}}
\def\p{\phi}
\def\F{\Phi}
\def\fc{\phi_{\rm c}}
\def\f0{\phi_0(r,t)}
\def\df{\delta\phi(r,t)}
\def\s{\sigma}
\def\a{\alpha}
\def\d{\delta}
\def\t{\tau}
\def\r{\rho}
\def\beq{\begin{equation}}
\def\eeq{\end{equation}}
\def\ba{\begin{eqnarray}}
\def\ea{\end{eqnarray}}
\def\re#1{[{\ref{#1}]}}

\vskip 1cm
\vspace{16pt}

A remarkable consequence of nonlinear field
theories is the existence of localized, non-singular solutions of
the classical equations of motion which are
non-dissipative. In general, these solutions can be time-dependent or
static.
As is well-known, the existence and simplicity of static
solutions is severely
constrained by dimensionality \re{DERR}. For a single self-interacting real
scalar field $\p (\x,t)$ such solutions are only
possible in (1+1)-dimensions.
More
realistic (3+1)-dimensional static solutions must invoke
more than one field,
as in the case of the 't Hooft-Polyakov monopole \re{RAJ}.

Given their relevance to the study of nonperturbative effects in
field theories, static, non-dissipative  solutions have been, for the
last twenty years or so,
the focus of most efforts in the study of nonlinear
solutions in classical field
theories. However, within the last decade, the possibility that spontaneous
symmetry breaking occurred in the early Universe has called for a better
understanding of time-dependent phenomena in the context of relativistic
field theories. For example, the dynamics of cosmological
phase transitions \re{KOLB} naturally
invokes out-of-equilibrium conditions, with fields interacting with
themselves and with a hot plasma in the background of an
expanding Universe \re{TDFT}.

In the present work, the possibility that {\it time-dependent}, localized,
non-dissipative solutions exist in the context of simple (3+1)-dimensional
scalar field theories is examined. In particular, the focus will be on
models involving only a single real scalar field with self-interactions
dictated by a double-well potential.
Since
it is known that for a symmetric double-well potential (SDWP)
all field configurations are unstable,
it is possible to obtain the lifetime of a given spherically
symmetric field configuration
by numerically evolving the equation of motion. By addopting this procedure,
it has been shown in the mid-seventies that certain configurations evolved
into a state which was considerably long-lived (even though the lifetimes
measured then were not very accurate). These time-dependent solutions
were called ``pulsons'' \re{PULSON}.
However, not much has been done in order
to further explore the properties of pulsons.
A few exceptions, which are
mainly related to the existence of these solutions for
the Sine-Gordon potential, different symmetries,
and somewhat contrived stability studies, are listed in
Ref.~ \ref{PULSONII}.
In fact, recent studies of bubble evolution in  SDWPs
overlooked the existence of pulsons \re{WIDROW}. Also, their
existence has never
been investigated for ADWPs.

Here, it will be argued that the existence of pulsons
is a very general feature
of models with both symmetric {\it and} asymmetric potentials,
depending only on
the initial amplitude and energy of the configuration. It will always be
assumed that the initial configuration interpolates between the two minima
of the potential (bubbles with amplitude below a
certain value cannot evolve
into pulsons, as will be clear later),
and that it is smooth
enough, say, $\p(r,t=0) \sim {\rm exp}[-r^2/R_0^2]$, or $\sim {\rm tanh}
(r-R_0)$, with $R_0$ the initial ``radius''. (From now on these
configurations will be called the gaussian and the tanh-bubbles.)
It is remarkable that
during the nonlinear evolution of these configurations a regime of
dynamical stability may be achieved in which energy is
practically conserved
within a localized region, due to some as yet unclear
conspiracy between the
many degrees of freedom involved.

The action for a real scalar field is,
$S[\p]=\int d^4x \left [{1\over 2}\partial_{\mu}
\p\partial^{\mu}\p - V_{{\rm S
(A)}}(\p)
\right] $,
where the subscripts S and A stand for the SDWP and ADWPs, with,
$V_{{\rm S}}(\p)={{\l}\over 4}\left (\p^2 -{{m^2}\over {\l}}\right )^2,
$ and
$V_{{\rm A}}(\p)={{m^2}\over 2}\p^2 - {{\a}\over 3}m\p^3 +
{{\l}\over 4}\p^4$.
A field configuration $\phi_0({\bf x},t)$ is a solution
of the equation of motion (an
overdot denotes partial time derivative),
$~{\ddot \p} - \nabla^2\p = -{{\partial V(\p)}\over {\partial \p}}$,
and has an energy,
\beq
E\left [\p_0\right ]=\int d^3x\left [{1\over 2}{\dot \p_0}^2+{1\over 2}
\left(\nabla\p_0\right )^2 + V(\p_0)\right ] \:.
\label{e:energy}
\eeq

For a sufficiently well-localized spherically symmetric
configuration $\f0$ (the case of interest here) with linear ``size''
$\sim R_0$, the integral over all
space can be restricted to a spherical volume containing the configuration,
\beq
E\left [\p_0\right ](t)=E_K(t) + E_S(t) + E_V(t)\:,
\label{e:ep}
\eeq
where  the kinetic, surface, and volume energies are defined respectively by
($\epsilon \ge 2$)
\beq
E_K(t)=2\pi\int_0^{\epsilon R_0}dr~r^2~{\dot \phi_0}^2,
{}~~E_S(t)=2\pi\int_0^{\epsilon R_0}dr~r^2~
\left (\phi_0^{\prime}\right )^2,~~E_V(t)=4\pi\int_0^{\epsilon
R_0}dr~r^2~V(\p_0) \:.
\label{e:ek}
\eeq

Introducing the dimensionless variables
$\F(r,t)=\sqrt{\l}\p(r,t)/m$, $\r=m~r$, and $\t=m~t$, the equation of motion
becomes, for both SDWP (and ADWP),
\beq
{{\partial^2\F}\over {\partial\t^2}}-{{\partial^2\F}\over {\partial \r^2}}
-{2\over {\r}}{{\partial \F}\over {\partial \r}}  =  \F - \F^3
{}~~~(-\F + {\tilde \a}\F^2 -\F^3 )\:,
\eeq
where ${\tilde \a}\equiv \a/\sqrt{\l}$. Note that for the ADWP,
a solution where $\F=0$ is the local minimum is possible
only if ${\tilde\a}^2
> 9/2$. The equation above was solved numerically
using a finite differencing
method fourth order accurate in space and second order accurate in
time. Since the problem is two-dimensional a very fine grid could be
used. By taking the lattice spacing to be $h=10^{-2}$ and the time step
to be $\theta=5\times 10^{-3}$, energy was conserved throughout
the evolution
to one part in $10^5$.

Let us concentrate on the SDWP for now. In this
case all bubbles are unstable, since there is no gain in volume energy
in going from one vacuum to another. Take the vacuum to be at $\F=-1$ and
consider configurations which interpolate between the two minima.
Both thick and thin-wall bubbles will be considered.
In Fig. 1 the energy within a spherical shell, Eq.~\ref{e:ep},
is shown for initial configurations, $\F_0(\r,0)=-{\rm tanh}(\r-\r_0)$, with
$\r_0=3$ and $\r_0=10$, and for $\F_0(\r,0)=2{\rm exp}\left [-\r^2/\r_0^2
\right ]-1$,
with
$\r_0=4$ and $\r_0=8$. Note the existence of an extended
period of stability,
of duration $\sim 10^3-10^4 m^{-1}$,
where practically no energy is radiated away. Thus, a necessary condition
(which was verified numerically) for
the existence of this pseudo-stable behavior is, from Eq.~\ref{e:ep}
\beq
{{dE_K}\over {dt}} \simeq -\left
({{dE_S}\over {dt}} +{{dE_V}\over {dt}}\right ) \:.
\label{e:cond}
\eeq

An interesting point
is that the value of the energy at the plateau is fairly independent
of the initial configuration. This suggests
the existence of an attractor-like
configuration in field space which is approached in the course of the
bubble's evolution.
An extensive (but not exaustive)
search indicates that only for configurations of initial energies
$E_0\ga 60m/\l\equiv E_{{\rm crit}}$ this behavior occurs.
This corresponds to a
gaussian bubble of radius $R_0\sim 2.4/m$. For smaller energies, the
evolution is well-fitted by the relation
$E(t)=E_0{\rm exp}[-t/\tau_L(R_0)]$,
where $\tau_L(R_0)$ is the bubble's lifetime. For gaussian bubbles
of radii $mR_0=1,~2$, $m\tau_L(R_0)=3.5,~12$, respectively.

{}From Fig.~1 it is clear that the evolution of large enough bubbles can be
divided into three stages. First the bubble sheds its initial energy by
quickly shrinking into a thick-wall bubble of energy roughly $E\sim 50m/\l$.
For thin wall bubbles, like the tanh-bubble with $\r_0\gg 1$,
this shrinking is well
described by the relativistic motion of the bubble wall. (See, e.g., Ref.
{}~\ref{WIDROW}.) The field then settles into the
pulson configuration. At any
given time, the configuration is well-approximated
by a half-gaussian, but with
softer asymptotic behavior, $\F(\r\rightarrow\infty ,\t)
\sim {\rm exp}(-\r)$.
The field is localized within a small volume with linear size
$\sim 3/m$, while its amplitude is rapidly
oscillating about a value between
the two minima.
(In this author's opinion, a better name for
these configurations would be
``oscillons''.)
Finally, during the last stage of the evolution,
the amplitude of oscillations decreases and the pulson quickly
radiates its remaining energy away.

In order to gain some insight into the properties of the pulson consider
the behavior of radial perturbations about $\F_0(\r,\t)$, defined as
$\F(\r,\t) = \F_0(\r,\t) +\d\F(\r,\t) $.
Since $\F_0(\r,\t)$ satisfies the equation of motion, expanding
$\d\F(\r,\t)$ in normal modes,
$\d\F (\r,\t)={\rm Re}\sum_n\p_n(\r){\rm exp}(i\omega_n \t)$,
it is found that
the amplitudes $\p_n(\r)$ satisfy the radial Schr\"odinger equation,
(only the $\ell=0$ mode will be considered here),
\beq
-{{d^2\p_n}\over {d\r^2}}- {2\over {\r}}{{d\p_n}\over {d\r}}
+V(\r,\t_0)\p_n = \l_n\p_n \:,
\label{e:schr1}
\eeq
where $V(\r,\t_0)\equiv 3\F_0^2(\r,\t_0) - 3$, and
$\l_n\equiv \omega_n^2 - 2$.
$\l_n$ is introduced to make sure that $V(\r\rightarrow
\infty)\rightarrow 0$. The stability of a given configuration  at
time $\t_0$, $\F_0(\r,\t_0)$, is determined by the lowest eigenvalue being
positive, $\omega_0^2 > 0$, or $\l_0 > -2$.

Clearly the general problem is quite complicated, as the potential
$V(\r,\t)$ is
time-dependent. A reasonable simplification is to solve this equation for a
given time $\t_0$ for which the field configuration $\F_0(\r,\t_0)$ is
obtained by evolving the equation of motion. This implicitly assumes
that $\F_0(\r,\t)$ varies slower than $\d\F(\r,\t)$. Otherwise, unstable
modes would not have enough time to grow and destabilize the pulson.
One can then find the eigenvalue
for a succession of snapshots and thus investigate its time evolution.
The Schr\"odinger equation was solved using the shooting method
\re{PRESS}.
To make
sure the method worked, the equation was solved for the Coulomb potential
and for the ``kink potential'' $ V(\r)=3{\rm cosh}^{-2}(\r/\sqrt{2})$,
since in both cases the eigenvalues are known analytically \re{LANDAU}.
(For the
kink case, one must recall that
the first eigenvalue in 3d corresponds to the second
in 1d, $E_0(3d)=E_1(1d)=-1/2$, due to the boundary
conditions at the origin.)

The results are shown in Fig. 2 during part of the pulson stage. Clearly,
the pulson is stable against small radial perturbations. (No instability
was detected during the whole pulson stage using the above method. An
argument to explain the pulson's final disappearance is advanced shortly.)

Next, the existence of pseudo-stable behavior for ADWPs
is investigated. The
asymmetry is controlled by the dimensionless
coupling ${\tilde \a}$. As two
examples, consider the nearly degenerate case, ${\tilde \a}=2.16$,
and the non-degerate case, ${\tilde \a}=2.23$.
In the context of first-order
phase transitions (finite temperature), a state initially localized at
$\F=0$ will decay by the nucleation of bubbles
larger than the critical bubble,
which is an extremum of the $O(3)$-invariant Euclidean action. The critical
bubble is thus the energy barrier for vacuum decay. For a given value of
${\tilde \a}$ it is easy to obtain the critical bubble and its
energy numerically,
by using the shooting method. (Now, the method finds the value of the field
in the bubble's core.) For the two values of the asymmetry above, the
energies and radii of the critical bubble are, respectively,
$E({\tilde \a}=2.16)\simeq  6.12\times 10^2m/\l$,
$R({\tilde \a}=2.16)\simeq
19.6/m$, and
$E({\tilde \a}=2.23)\simeq 1.17\times 10^2m/\l$,  $R({\tilde \a}=2.23)
\simeq 7.9/m$.
The interest here is in investigating the evolution
of {\it sub-critical} bubbles, to see if they display the
pseudo-stable behavior
observed for the SDWP. The results are shown in Fig. 3 where the energy
within a spherical shell surrounding the initial configurations
is shown as a
function of time.
For each value of ${\tilde \a}$, two tanh-bubbles with
initial radii $\r_0=3$
and $\r_0=4$ were examined.
Again, the existence of very long-lived pulsons is observed,
with $E_{{\rm
crit}}$ and lifetimes depending on the asymmetry.
In Fig. 4 the phase-space portrait of the pulson's core ($\r=0$), for
${\tilde \a}=2.16$ and $\r_0=4$
is displayed for $\t\ge 1000$. Note the similarity with the motion of
a ``damped'' (an)harmonic oscillator. During the pulson stage,
the motion is
restricted to a band in phase-space.
As the
pulson becomes unstable, it will spiral around the final stable point
$\F=0$ and ${\dot \F} =0$; As energy is gradually radiated away,
the maximum possible amplitude (that is for ${\dot \F}\simeq 0$)
is driven below a critical value
for stability. Bubbles with initial amplitudes below this value,
roughly about the maximum of $V(\F)$,  {\it and} small kinetic energy
will quickly shrink.
Note that this is
the value below which the potential is approximately parabolic, so that
nonlinear effects
become subdominant.  (Of course, small amplitude but high velocity bubbles
can still escape the attractive well centered at $\F=0$, as during the
pseudo-stable pulson stage.)

The present work raises many questions for future investigation. Apart
from a more detailed study of the pulson's properties and stability,
it is still not clear why the initial configurations evolve into the pulson
stage. The pulson's behavior is suggestive of some
sort of nonlinear resonance effect occurring between the different
modes of the field. It should be possible to separate the field into short
and long wavelengths, with a cutoff around $m^{-1}$. The short wavelength
modes would act as a perturbation, which due to the nonlinear coupling might
induce the observed behavior.
Apart from their potential interest for, among other topics,
nonlinear optics and long Josephson
junctions \re{REVIEW}, these solutions are
also of interest in the context of
cosmological phase transitions \re{KOLB}.
Consider two possible regimes of interest, defined by the ratio of time
scales ${\cal R}=\t_{\phi}/\t_U\sim (\phi/{\dot \phi})/(\mpl/T^2)\sim
T/\mpl$, where $T$ is the temperature and $\mpl$ is the Planck mass.
For  ${\cal R}\sim 10^{-3}$, corresponding to the GUT scale,
the extended lifetime of one sub-critical bubble can be relevant
to the dynamics of a first-order transition.
The Universe cools off quite fast
and if the sub-critical bubble lives long enough
it can {\it become} a critical
bubble. For ${\cal R}\sim 10^{-16}$, corresponding to the electroweak
scale, many sub-critical bubbles can be present
before a critical bubble nucleates \re{MG}.
Given their long lifetimes, they could serve as nucleation sites for
critical bubbles, very much like impurities in ordinary phase transitions,
speeding the process of vacuum decay considerably. Finally, in the
context of late-time phase transitions, it is desirable to have bubbles
live longer in order for matter accretion to be efficient \re{OVRUT}. These
and related questions are currently under investigation.

\acknowledgements
I am grateful to J.D.  Harris for many enlightening comments and to R. Ramos
for useful discussions. I also thank M. Briggs for his numerical skills
during the early stages of this investigation.
This work is partially supported by a National Science Foundation grant
No. PHYS-9204726.

\references
\begin{enumerate}

\item\label{DERR} G. H. Derrick, J. Math. Phys. {\bf 5}, 1252 (1964).

\item\label{RAJ} R. Rajaraman, {\it Solitons and Instantons}, North-Holland
(1982).

\item\label{KOLB} E. W. Kolb and M. S. Turner, {\it The Early Universe},
Addison-Wesley (1990).

\item\label{TDFT} A few examples are, D. Boyanovsky and H. J. de Vega,
Phys. Rev. {\bf D47}, 2343 (1993); J. M. Cornwall and R.
Bruinsma, Phys. Rev. {\bf D38}, 3144 (1988); E. Calzetta and B. L. Hu,
Phys. Rev. {\bf D37}, 2878 (1988).

\item\label{PULSON} I. L. Bogolubsky and V. G. Makhankov, Pis'ma Zh.
Eksp. Teor. Fiz. {\bf 24}, 15 (1976) [JETP Lett.,
{\bf 24}, 12 (1976)]; {\it ibid.} {\bf 25}, 120 (1977) [{\it ibid.}
{\bf 25}, 107 (1977)]; V. G. Makhankov, Phys. Rep. {\bf C 35}, 1 (1978).

\item\label{PULSONII} I. L. Bogolubsky, Phys. Lett. {\bf A61}, 205 (1977);
J. Geicke, Phys. Scripta {\bf 29}, 431 (1984); B. A. Malomed and
E. M. Maslov, Phys. Lett. {\bf A160}, 233 (1991); A. M. Srivastava, Phys.
Rev. {\bf D46}, 1353 (1992).

\item\label{WIDROW} L. M. Widrow, Phys. Rev. {\bf D40}, 1002 (1989).

\item\label{PRESS} W. H. Press, B. P. Flannery, S. A. Teukolsky, and W. T.
Vetterling, {\it Numerical Recipes: The Art of Scientific Computing},
Cambridge University Press (1986).

\item\label{LANDAU} L. D. Landau and E. M. Lifshitz, {\it Quantum Mechanics;
nonrelativistic theory}, Addison-Wesley (1965).

\item\label{REVIEW} For a review see, {\it Solitons}, Eds. S. E. Trullinger,
V. E. Zakharov, and V. L. Prokovsky, North-Holland (1986).

\item\label{MG} M. Gleiser and E. W. Kolb, Phys. Rev. Lett. {\bf 69},
1304 (1992).

\item\label{OVRUT} See, for example, Z. Lalak and B. A. Ovrut, University
of Pennsylvania Report No. UPR-0504T; C. Hill, D. Schramm, and J. Fry,
Comm. on Nucl. and Part. Phys. {\bf 19}, 25 (1989).

\end{enumerate}

\listoffigures

Figure 1. The energy within a spherical shell surronding several initial
configurations as a function of time for the SDWP. Larger bubbles have
shorter lifetimes.\\

Figure 2. Time evolution of lowest eigenvalue. For clarity, points
with $\omega_0^2 >
2$ were displayed as $\omega_0^2=2$.\\

Figure 3.  The energy within a spherical shell surronding several initial
configurations as a function of time for the ADWP. \\

Figure 4. Phase-space evolution of pulson's core for ${\tilde \a}=2.16$,
$\r_0=4$,
and $\t\ge 1000$.

\end{document}